\documentclass[fleqn,twoside]{article}
\usepackage[headings]{espcrc2}
\usepackage{subfigure}
\usepackage{graphicx}

\title{On the measurement of the proton-air cross section using longitudinal shower profiles}
\author{R.~Ulrich\address[FZK]{Forschungszentrum Karlsruhe, Institut f\"ur Kernphysik, Germany}, J.~Bl\"umer\addressmark, R.~Engel\addressmark, F.~Sch\"ussler\addressmark, M.~Unger\addressmark}
\runtitle{On the measurement of $\sigma_{\rm p-air}$ using longitudinal shower profiles}
\runauthor{Ralf~Ulrich et al.}

\begin{document}

\begin{abstract}
In this paper, we will discuss the prospects of deducing the proton-air cross section from 
fluorescence telescope measurements of extensive air showers.
As it is not possible to observe the point of first interaction $X_{\rm 1}$ directly, 
other observables closely linked to $X_{\rm 1}$ must be inferred from the longitudinal 
profiles. This introduces a dependence on the models used to describe the shower 
development. The most straightforward candidate for a good correlation 
to $X_{\rm 1}$ is the depth of shower maximum $X_{\rm max}$.
We will discuss the sensitivity of an $X_{\rm max}$-based analysis on $\sigma_{\rm p-air}$
and quantify the systematic uncertainties arising from the model dependence, 
parameters of the reconstruction method itself and a possible non-proton contamination 
of the selected shower sample.
\end{abstract}

\maketitle

\section{Introduction}
One of the most important parameter needed for interpretation of air
shower measurements is the proton-air cross section $\sigma_{\rm
  p-air}$, which is not well known at cosmic ray energies.
Collider measurements of the proton-(anti)proton cross section have to
be extrapolated over a wide energy range to be used in models of
nuclear effects to predict the proton-air cross section at ultra high 
energies. On the other
hand, it might be possible to measure $\sigma_{\rm p-air}$ by
analyzing extensive air showers (EAS).

The point of first interaction in slant depth $X_{\rm 1}$ follows the
exponential distribution 
\begin{equation}
  \frac{dP(X_{\rm 1})}{dX_{\rm 1}}=\frac{1}{\lambda_{\rm p-air}} \cdot e^{-X_{\rm 1}/\lambda_{\rm p-air}} \textrm{.}
\end{equation}
The cross section $\sigma_{\rm p-air}$ is related to the absorption length 
$\lambda_{\rm p-air}$ as $\lambda_{\rm p-air}=M_{\rm mean}/\sigma_{\rm p-air}$,
with $M_{\rm mean}$ being the mean mass of a target (air) particle.

The depth and characteristics of the first interaction of the primary cosmic ray particle
with a molecule of the atmosphere 
determines to a large extent the resulting shower shape. This 
implies strong correlations of the first interaction with experimentally 
observable parameters of the shower. 

Several air shower experiments have provided measured values of $\sigma_{\rm p-air}$ by 
exploiting correlations of $X_{\rm 1}$ with observables found by using air shower simulations 
\cite{Honda93,Hara83,Baltrusaitis84,Aglietta97,Knurenko99}.
Of course, these results depend strongly on the used simulations and especially on 
the chosen high energy hadronic interaction model \cite{Gaisser87,Engel98}. 

\begin{figure*}[tb!]
\centerline{ 
\includegraphics[width=.499\linewidth]{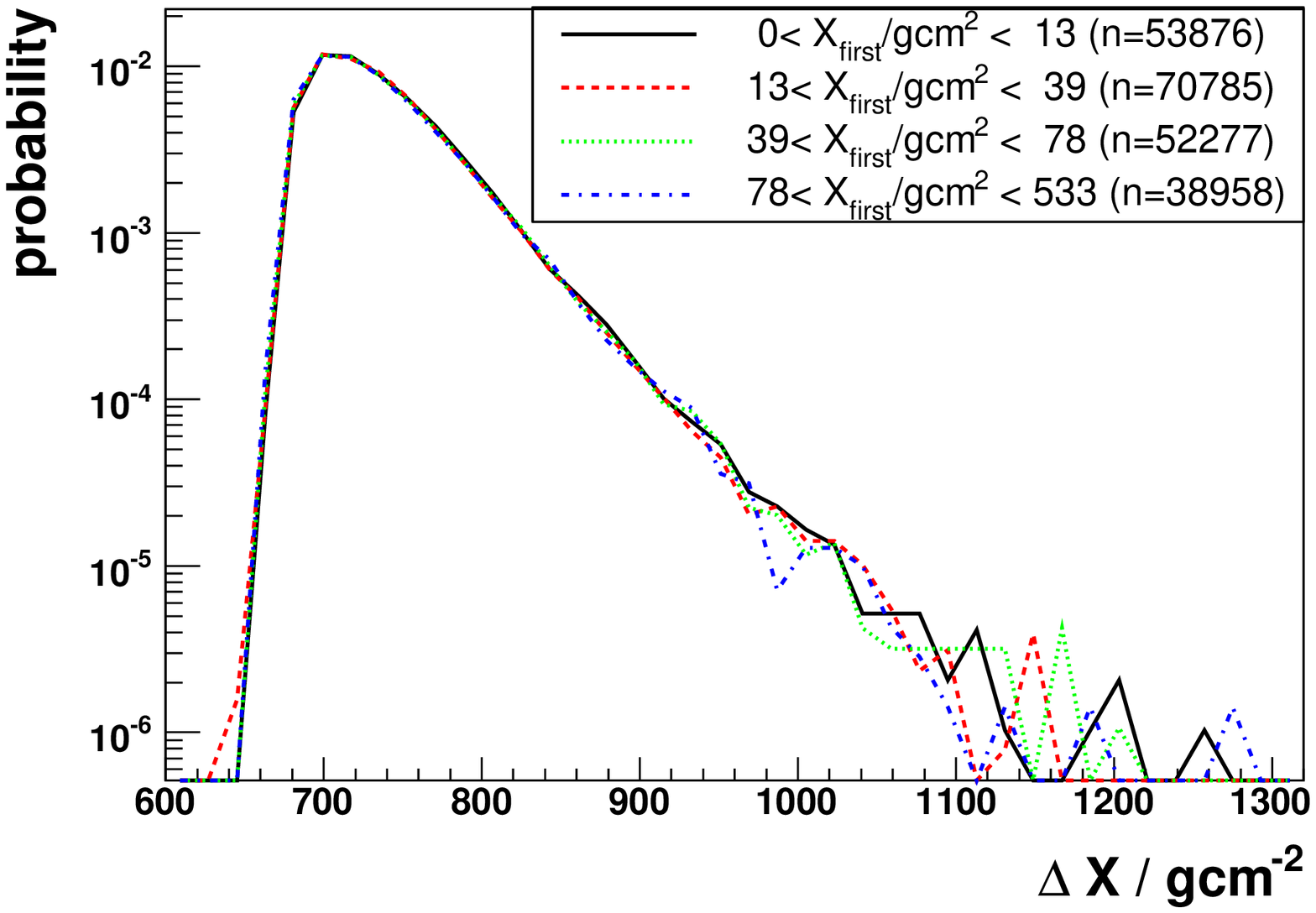}
\includegraphics[width=.499\linewidth]{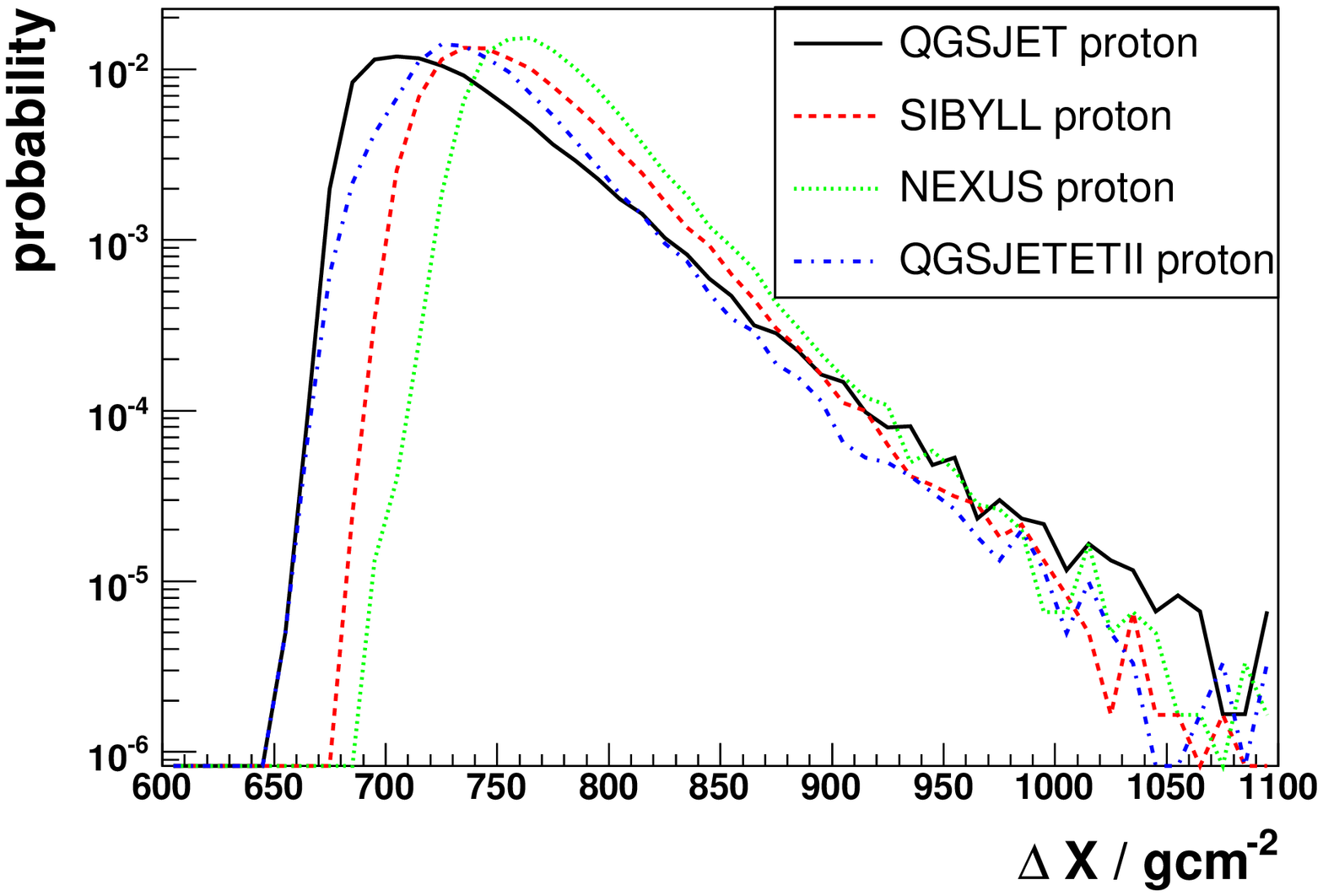}
}
\vspace*{-.5cm}
\caption{ Left panel: Dependence of the $\Delta X$-distribution on
  $X_{\rm 1}$.  
Right panel: Interaction model dependence of $\Delta X$-distribution.
  \label{f:DeltaX}
}
\end{figure*}

The HiRes collaboration proposed a new technique which appeared to be almost
independent of the hadronic interaction model \cite{Belov}. The idea is to 
decompose the measured distribution of shower maxima $dP(X_{\rm max})/dX_{\rm max}$ into 
the two independent distributions of $X_{\rm 1}$
and $\Delta X$ with
\begin{equation}
  X_{\rm max} = X_{\rm 1} + \Delta X .
\end{equation}

Using the introduced $\Delta X$ one can express the $X_{\rm max}$ distribution 
as folding integral
\begin{equation}
  P(X_{\rm max}) = \int_0^\infty \, dX_{\rm 1} \; P(X_{\rm 1}) \cdot P(\Delta X | X_{\rm 1}) 
\label{eq:folding}
\end{equation}
of the $X_{\rm 1}$ and $\Delta X$ distributions, whereas in the HiRes \cite{Belov} approach 
$P(\Delta X | X_{\rm 1})=P(\Delta X)$ is used.

Measuring $\sigma_{\rm p-air}$ means 
to deconvolve the $X_{\rm max}$ distribution using a known $\Delta X$ distribution.

The distribution of $\Delta X$ needs to be inferred from simulations.

\section{Simulation study}
  
To investigate the deconvolution technique, we simulated sets of 
proton-induced showers of 10~EeV using CONEX \cite{CONEX} with the high
energy interaction models QGSJET01 \cite{qgsjet01}, SIBYLL2.1
\cite{sibyll}, NEXUS3 \cite{nexus}, and QGSJETII.3 \cite{qgsjetII}.
In addition helium, CNO and gamma-ray primaries were simulated with
QGSJET01. Each set of simulations contains 200.000 events.

\section{Universality of $\Delta X$ distribution}

One of the important conditions for the applicability of the method
proposed in \cite{Belov} is the $X_1$-independence of the $\Delta X$
distribution. For a given model, this is indeed the case as shown in
Fig.~\ref{f:DeltaX} (left). The universality in $X_{\rm 1}$ has been
verified for proton primaries using NEXUS3, SIBYLL2.1, QGSJETII.3, and
QGSJET01 at 1~EeV and 10~EeV.

Therefore one can use
\begin{equation}
  P(\Delta X | X_{\rm 1}) = P(\Delta X) .
\end{equation}

However, the resulting $\Delta X$ distribution depends strongly on the hadronic interaction model
used for shower simulation (Fig. \ref{f:DeltaX}, right). The maximum of the $\Delta X$ distribution is shifted by
up to 60~g/cm$^2$ and the slope beyond the maximum changes by up to a factor of 1.36. 
This introduces a principal model dependence to the deconvolution technique. It can be tried
to reduce this dependence by adding an additional degree of freedom, $X_{\rm shift}$,
which allows the $\Delta X$ distribution to shift along the x axis
\begin{equation}
  P(\Delta X) \rightarrow P(\Delta X + X_{\rm shift}) .
\end{equation}
It is much more difficult to reduce the model-dependence of the shape of the $\Delta X$ 
distribution. This will not be discussed in this paper but we shall show the impact of this model-dependence 
on the resulting $\sigma_{\rm p-air}$ reconstruction.

\section{Study of the deconvolution method}
  
\begin{table*}[t!]
  \caption{
    Analysis results of $\sigma_{\rm p-air}$ in mb for {10~EeV} proton showers. The results for 
    $X_{\rm shift}$ are given in g/cm$^{-2}$. 
    The expected cross sections are 532\,mb (QGSJET01), 612\,mb (SIBYLL 2.1), 542\,mb (NEXUS), and 562\,mb (QGSJETII).
 \label{t:bigtable}
  }
  \scriptsize
  \centering
  \begin{tabular}{l||cc|cc|cc|cc}
    \hline
    \multicolumn{1}{c||}{data} & \multicolumn{8}{c}{$\Delta$ X model used for reconstruction}\\
    \hline
    & \multicolumn{2}{|c}{QGSJET} & \multicolumn{2}{|c}{SIBYLL} & \multicolumn{2}{|c}{NEXUS} & \multicolumn{2}{|c}{QGSJETII} \\
    data model & $\sigma_{\rm rec}$ & $X_{\rm shift}$ & $\sigma_{\rm rec}$ & $X_{\rm shift}$ & $\sigma_{\rm rec}$ & $X_{\rm shift}$ & $\sigma_{\rm rec}$ & $X_{\rm shift}$ \\
    \hline
    QGSJET01 & \bf 537.1$\pm$5.5 & 1.7$\pm$0.8  & 465.3$\pm$4.1     & -29.8$\pm$0.8 & 447.9$\pm$3.2     & -46.9$\pm$0.8 & 467.6$\pm$3.7     & -19.6$\pm$0.7 \\
    SIBYLL & 802.8$\pm$12      & 36.1$\pm$0.7 & \bf 609.2$\pm$5.7 & 0.7$\pm$0.5   & 572.8$\pm$4.5     & -17.4$\pm$0.5 & 613.3$\pm$5.7     &  10.6$\pm$0.5 \\
    NEXUS & 749.9$\pm$10      & 55.1$\pm$1.2 & 569.1$\pm$3.9     & 19.5$\pm$0.9  & \bf 543.1$\pm$4.0 & 1.5$\pm$0.5   & 576.5$\pm$4.6     & 29.5$\pm$0.8 \\
    QGSJETII & 697.1$\pm$8.2     & 27.0$\pm$0.8 & 553.2$\pm$4.3     & -7.5$\pm$0.6  & 521.6$\pm$3.7     & -26.1$\pm$0.3 & \bf 562.3$\pm$4.4 & 2.1$\pm$0.4 \\
    \hline
  \end{tabular}
\end{table*}

For each model, a Monte Carlo data set of 50.000 events was generated
and analyzed with the $\Delta X$ distributions found previously. No
detector effects are considered. Two free parameters ($\lambda_{\rm
  p-air}$ and $X_{\rm shift}$) are fitted using $\chi^2$
minimization and
\begin{eqnarray}
  \frac{dP(X_{\rm max})}{dX_{\rm max}} & = & \int_{X_{\rm 1,min}}^{X_{\rm 1,max}} \, dX_{\rm 1} \; 
  \frac{1}{\lambda_{\rm p-air}} \; e^{-X_{\rm 1}/\lambda_{\rm p-air}} \nonumber \\
  & \times & \left( \frac{dP(\Delta X +X_{\rm shift})}{d\Delta X} \right)_{\textrm{\scriptsize model}}\;.
\end{eqnarray}

The fit results are shown in Tab.~\ref{t:bigtable} for all combinations of hadronic 
interaction models and $\Delta X$ distributions. The diagonal elements of the table 
represent matching models for data and $\Delta X$. In these cases the true cross section of 
the input model can be retrieved very well by the convolution method. It is interesting 
that for models with a similar shape of the $\Delta X$ distribution
(QGSJETII and SIBYLL) the reconstruction still works rather
well, whereas for combinations of models which have a very different shape of $\Delta X$ 
(QGSJET and NEXUS) the results differ up to 200~mb.

\subsection{Importance of  fit range}

\begin{figure*}[t!]
  \centerline{
    \includegraphics[width=.49\linewidth]{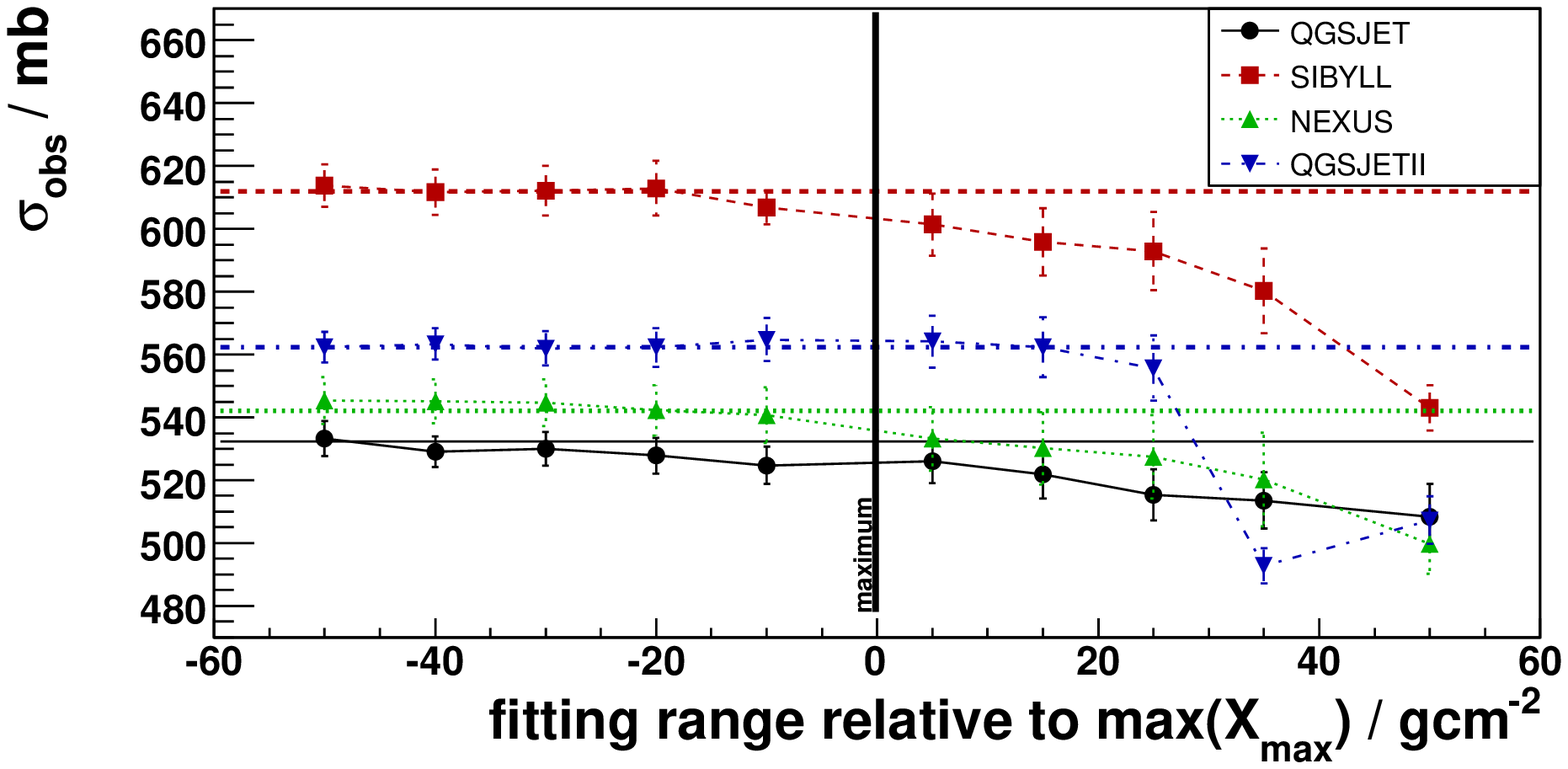}~
    \includegraphics[width=.49\linewidth]{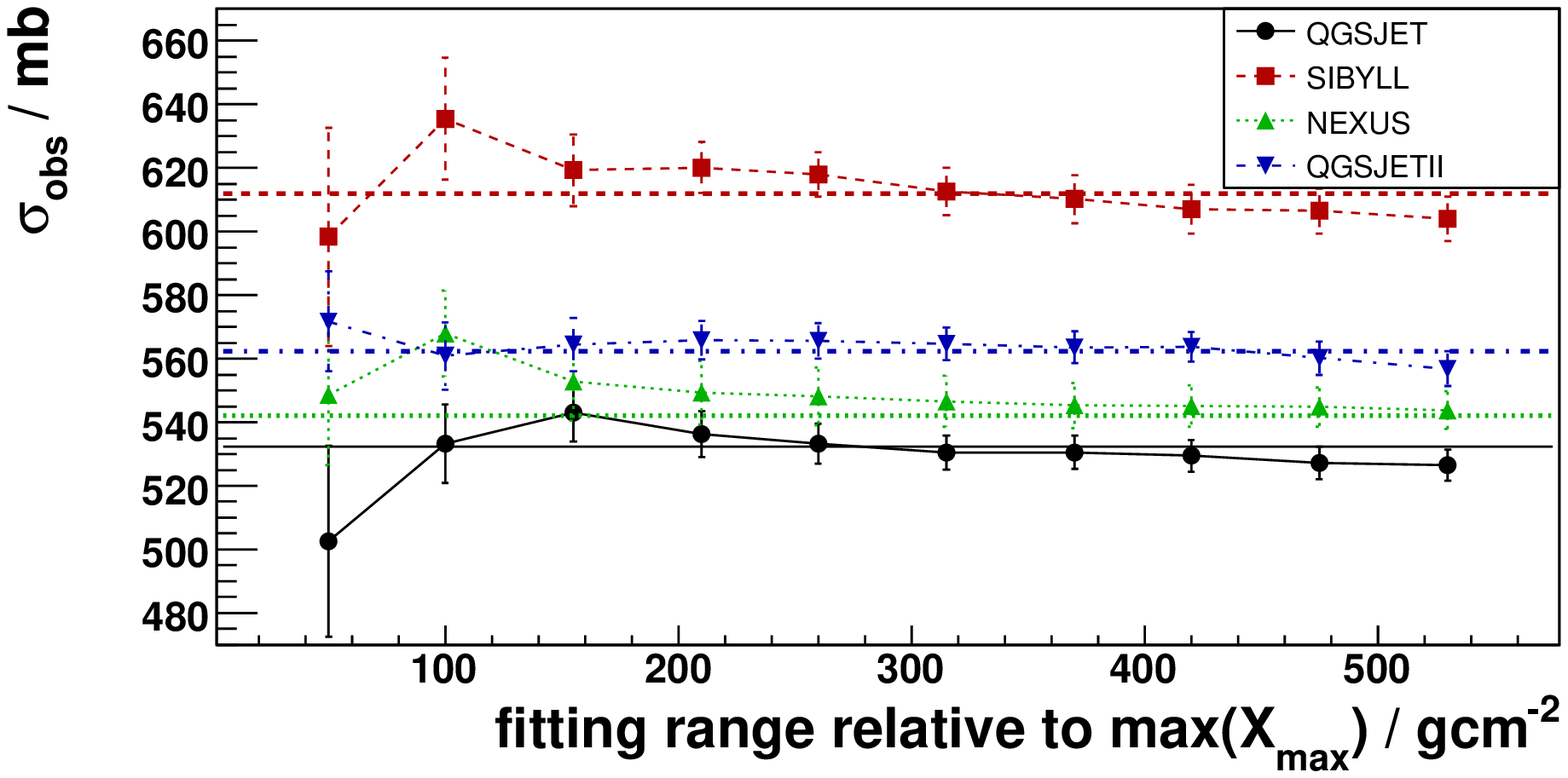}
  }
  \vspace*{-.5cm}
  \caption{Impact of the begin (left panel) and the end (right panel)
    of the fit range of the $X_{\rm max}$ distribution (pure proton composition at 10~EeV). 
    On the x axis the distance to the peak of the $X_{\rm max}$ distribution is shown. The
    horizontal lines denote the true value of $\sigma^{\rm true,\; model}_{\rm p-air}$ for each interaction model.
  \label{f:fittingRange}
}
\end{figure*}

The reconstructed value of $\sigma_{\rm p-air}$ is
very sensitive to the chosen starting point of the fit range, as shown in
Fig.~\ref{f:fittingRange} (left).
The reason for this is the introduced $X_{\rm shift}$
parameter which needs the peak of the $X_{\rm max}$ distribution to be restricted. 
To fit only the slope after the peak of the $X_{\rm max}$ distribution is not 
sufficient if $X_{\rm shift}$ is treated as free parameter.

Restricting the end point of the fitting range mainly increases the
statistical uncertainty of the fit result. The true value of
$\sigma^{\rm true}_{\rm p-air}$ is still within the
statistical errors of the reconstructed values (see Fig.~\ref{f:fittingRange}, right).

\subsection{Impact of other elements}

\begin{figure}[b!]
  \begin{center}
    \includegraphics[width=\linewidth]{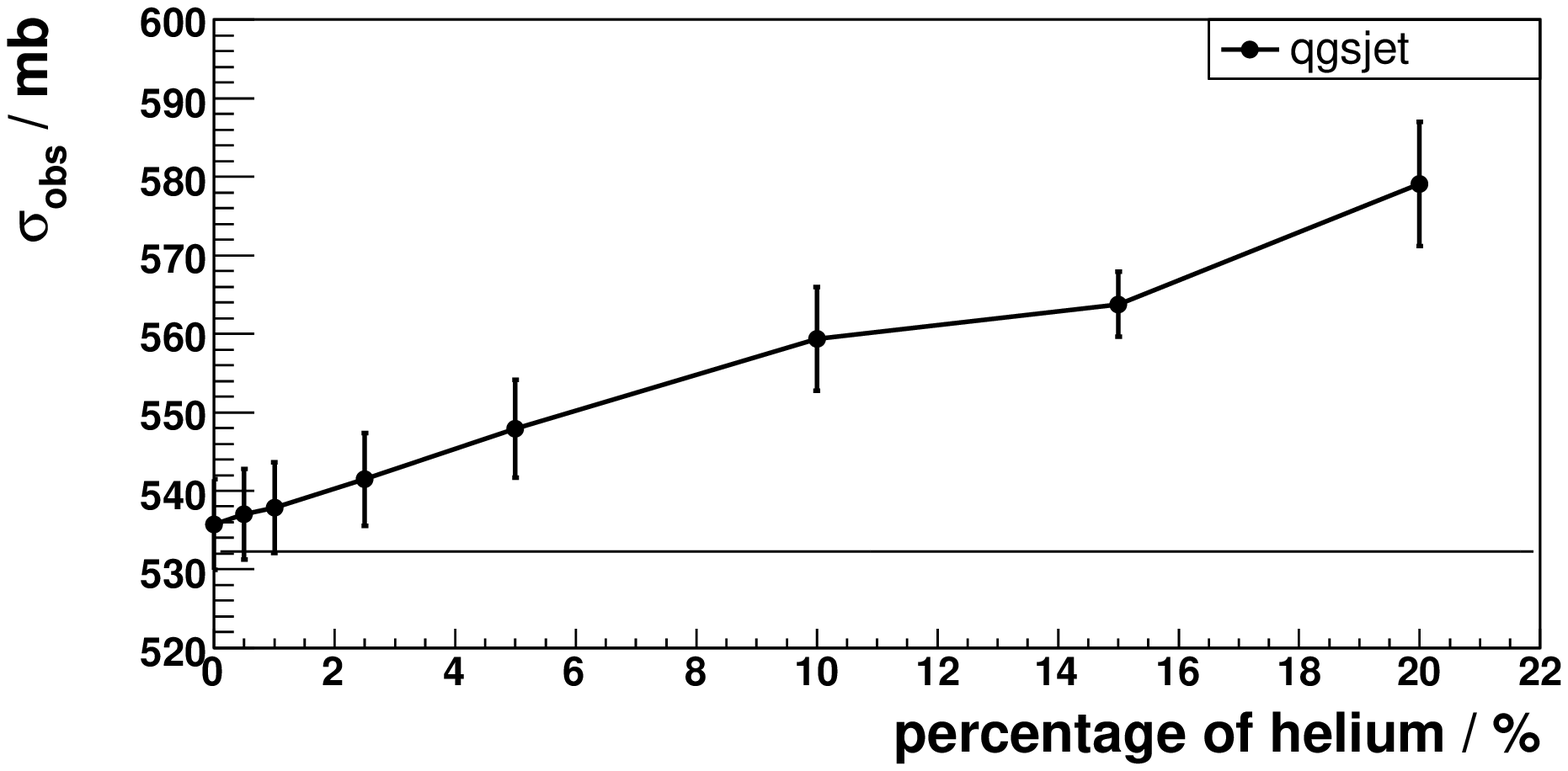}\\
    \includegraphics[width=\linewidth]{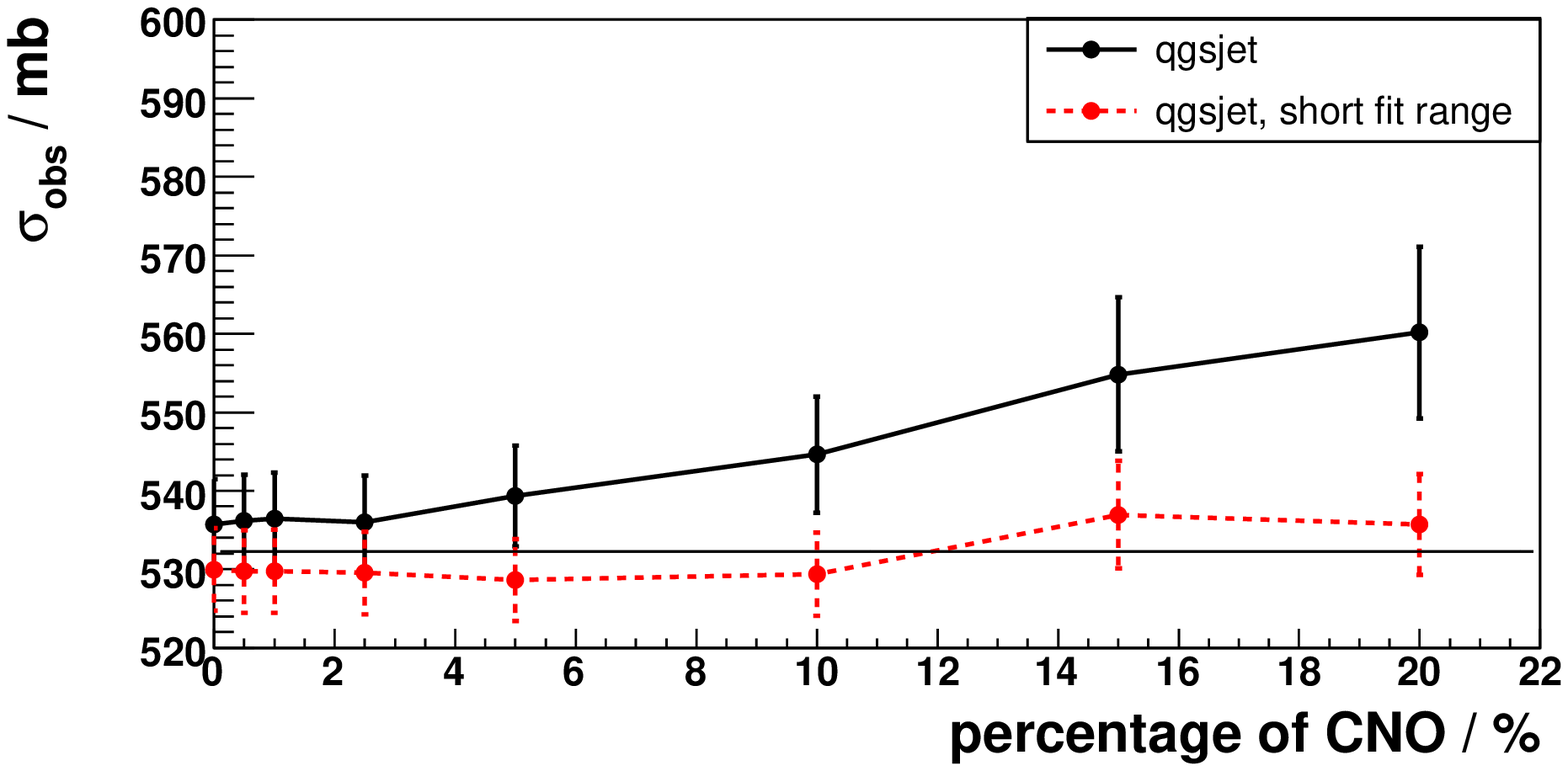}\\
    \includegraphics[width=\linewidth]{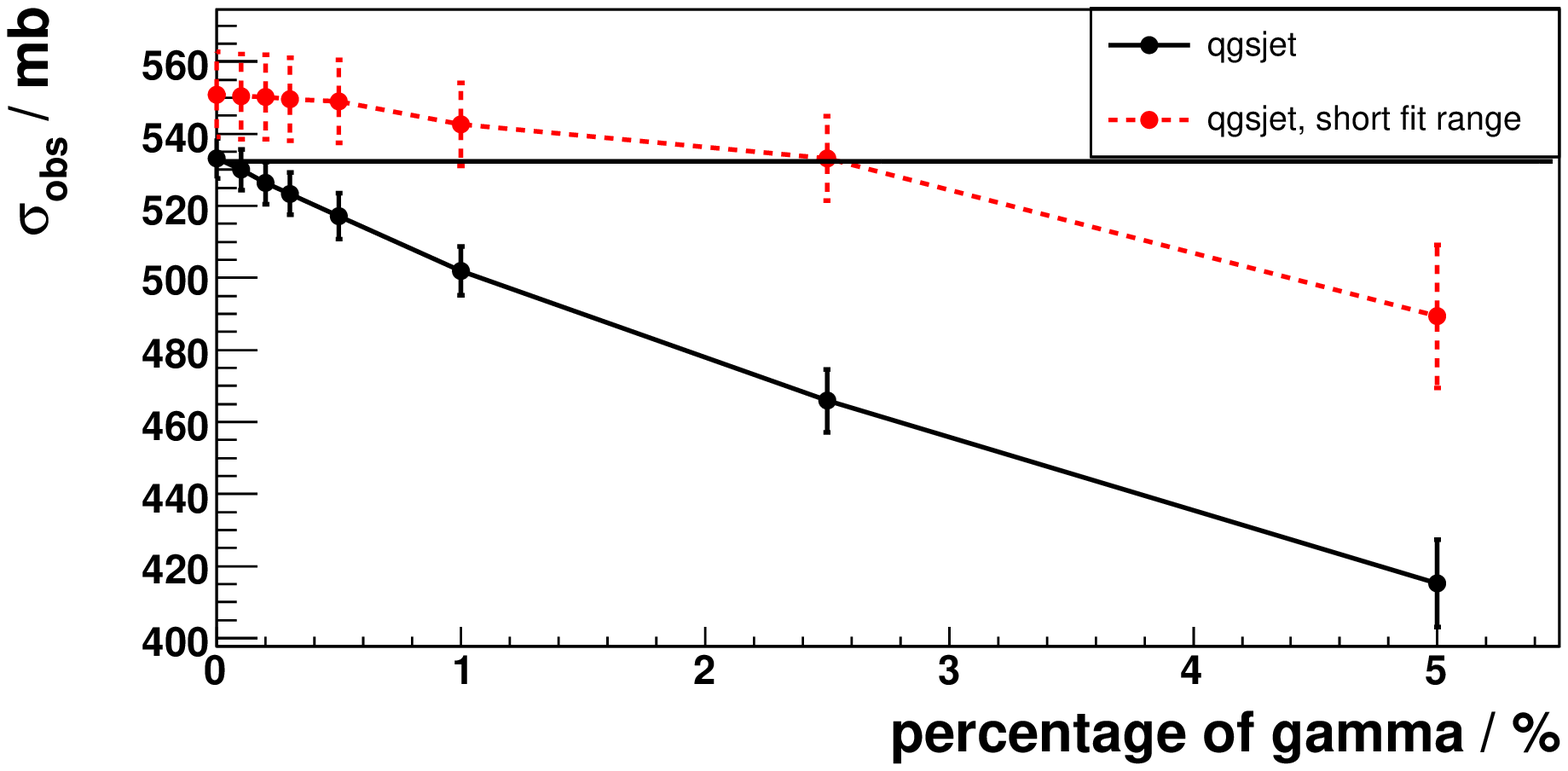}
  \end{center}
  \vspace*{-1cm}
  \caption{ Impact of non-proton contributions in the analyzed shower
    sets on the reconstruction of $\sigma_{\rm p-air}$. The horizontal
    lines denote the true value of $\sigma^{\rm true,\; model}_{\rm
      p-air}$ of the used high energy interaction model.
  \label{f:mix}
}
\end{figure}

If the selected showers do contain a fraction of light elements other than protons, the shape of the 
measured $X_{\rm max}$ distribution will be altered. 
A possible contribution of
helium will mostly distort the region of the peak of the $X_{\rm max}$ distribution.
Since the peak of the $X_{\rm max}$ distribution
is important for the fit itself, the reconstruction of $\sigma_{\rm p-air}$ is affected (Fig.~\ref{f:mix}, top)

Contamination with CNO showers mainly 
influences the rising edge of the $X_{\rm max}$ distribution. The impact on
the fit result can be limited by shifting the starting point of the fit range
more towards the peak of the distribution. Using a short fit range, the effect on the 
reconstruction of $\sigma_{\rm p-air}$ can be neglected for small 
fractions of CNO (Fig. \ref{f:mix}, middle).

\subsection{Impact of gamma-ray primaries}

Gamma-ray primaries produce an $X_{\rm max}$ distribution which is
shifted to higher $X_{\rm max}$ by about 200~g/cm$^2$ with respect to
protons. Therefore the influence of gamma-ray showers can be reduced
by setting the end of the fitting range appropriately
(Fig.~\ref{f:mix}, bottom). Nevertheless already a small contribution of
gamma-rays does affect the $X_{\rm max}$ distribution significantly
because of the low number of proton events at high $X_{\rm
  max}$. 
It should be noted that even if cosmic ray sources would accelerate
only protons, a small fraction of gamma-rays is expected due to the
GZK energy loss process \cite{Gelmini}.

\section{Summary}

The folding integral method would be very powerful given the 
true $\Delta X$ distribution is known. Input cross sections can 
be reconstructed within statistical uncertainties of 
a few mb. Unfortunately, the $\Delta X$ distribution does depend on 
the high energy hadronic interaction model used during air shower 
simulation. The $\Delta X$ model
dependence can be reduced by adding a new degree of 
freedom $X_{\rm shift}$ to the fit but it still amounts 
up to $\sim$~100~mb. On the other hand, treating $X_{\rm shift}$ as free parameter 
increases the sensitivity
of $\sigma_{\rm p-air}$ to a possible helium contamination of the 
$X_{\rm max}$ distribution. Medium mass cosmic ray primaries
have only limited influence on the reconstructed $\sigma_{\rm p-air}$. 
Even a small fraction of
gamma-ray primaries has a significant impact on the 
reconstruction results. 

To use the folding integral method for a reliable cross section measurement, the
model dependence of the $\Delta X$ distribution has to be reduced
significantly.

\begin{thebibliography}{9}
\bibitem{Honda93} M. Honda et al., Phys. Rev. Lett. 70 (1993) 525
\bibitem{Hara83} T. Hara et al., Phys. Rev. Lett. 50 (1983) 2058
\bibitem{Baltrusaitis84} R. M. Baltrusaitis et al., Phys. Rev. Lett. 52 (1984) 1380 
\bibitem{Aglietta97} M. Aglietta et al. Nucl. Phys. A (Proc.Suppl.) 75A (1999) 222
\bibitem{Knurenko99} S. P. Knurenko et al., 26th ICRC Utah 1 (1999) 372 
\bibitem{Gaisser87} T. K. Gaisser et al., Phys. Rev. D 36 (1987) 1350
\bibitem{Engel98} R. Engel et al., Phys. Rev. D 58 (1998) 014019
\bibitem{Belov} K. Belov et al., Nucl. Phys. (Proc.Suppl.) 151 (2006) 197
\bibitem{CONEX} T. Bergmann et al., Astropart. Phys. 26 (2007) 420
\bibitem{qgsjet01} N.N. Kalmykov et al., Nucl. Phys. B (Proc. Suppl.) 52B (1997) 17
\bibitem{sibyll} R. Engel et al., Proc. 26th ICRC Utah 1 (1999) 415
\bibitem{nexus} K. Werner et al., Heavy Ion Phys. 21 (2004) 279
\bibitem{qgsjetII} S. Ostapchenko, Nucl. Phys. (Proc.Suppl.) 151 (2006) 143
\bibitem{Gelmini} G. Gelmini et al., astro-ph/0506128


\end{thebibliography}
\end{document}